\documentstyle[epsf]{article}

\def\setb@se#1{\baselineskip=#1 \normalbaselineskip=#1}
\setb@se{12pt}
\long\def\title#1{\vspace*{1pc}{\pretolerance=10000\raggedright
\setb@se{14pt}\large\bf #1\par}\nobreak\ignorespaces}
\long\def\author#1{\vspace{1pc}\begin{list}{\hfill}%
{\topsep=0pt\parskip=0pt\parsep=0pt\partopsep=0pt\listparindent=0pt%
\itemsep=0pt\rightmargin=0pt\labelsep=0pt\labelwidth=2pc\leftmargin=4pc}%
\item\normalsize{\rm #1}\end{list}\vspace{2pt}}
\long\def\affil#1{\begin{list}{\hfill}%
{\topsep=0pt\parskip=0pt\parsep=0pt\partopsep=0pt\listparindent=0pt%
\itemsep=0pt\rightmargin=0pt\labelsep=0pt\labelwidth=2pc\leftmargin=4pc}%
\item\small{\rm #1}\end{list}\vspace{2pt}}
\long\def\beginabstract{\vspace{12pt plus 4pt minus 2pt}\begin{list}{\hfill}%
{\topsep=0pt\parskip=0pt\parsep=0pt\partopsep=0pt\listparindent=0pt%
\itemsep=0pt\rightmargin=0pt\labelsep=0pt\labelwidth=2pc\leftmargin=0pc}%
\item\small{\bf Abstract. }}
\long\def\endabstract{\end{list}\vspace{1pc plus4pt minus 2pt}%
\normalsize\noindent}

\begin{document}
\title{Induced currents, frozen charges and the quantum Hall effect breakdown}
\author{K.V.~Kavokin\dag\ddag, M.E.~Portnoi\dag\ddag, A.J.~Matthews\dag,
J.~Gething\dag, A.~Usher\dag \\
D.A.~Ritchie\S, and M.Y.~Simmons\S}
\affil{\dag~School of Physics, University of Exeter, Exeter EX4 4QL, United
Kingdom\newline
\ddag~A.~F.~Ioffe Physico-Technical Institute, 194021 St~Petersburg, Russia
\newline
\S~Cavendish Laboratory, Cambridge CB3 0HE, United Kingdom}

\beginabstract
Puzzling results obtained from torque magnetometry in the quantum Hall 
effect (QHE) regime are presented, and a theory is proposed for their 
explanation.
Magnetic moment saturation, which is usually attributed to the QHE
breakdown, is shown to be related to the charge redistribution across 
the sample.  
\endabstract

Since the discovery of the quantum Hall effect (QHE), experiments using
strong magnetic fields and low temperatures have continued to give new
information on the physics of two-dimensional electron systems (2DESs).
However, much remains to be understood. One of the most puzzling phenomena is
the so-called QHE breakdown - a complex of nonlinear effects arising when
high-density currents are passed through a 2DES. Many theoretical models
have been proposed to explain it \cite{1}, although no universal agreement 
has yet been reached. Among the most fundamental reasons for QHE breakdown 
at some current density in any 2DES, the quantum
inter-Landau-level scattering (QUILLS)~\cite{2} deserves mention. 
The QUILLS process consists in tunneling of electrons between adjacent 
Landau levels (LLs),
which becomes possible when the in-plane electric field reaches values
comparable to $\hbar \omega _{c}/l_{H}$, where $\hbar \omega _{c}$ is the
cyclotron energy, and $l_{H}$ is the magnetic length. Much experimental
effort has been spent on detecting QUILLS, but no reliable results have been
so far obtained. There are many indications that the QHE breakdown observed
in traditional QHE experiments is strongly affected by contacts~\cite{1}. 
To avoid contact effects, it was proposed to induce currents in a 2DES 
by sweeping magnetic field and to detect these currents with a 
high-sensitivity torque magnetometer~\cite{3}. 
By using this technique, an effect
was observed which was qualitatively consistent with the QHE breakdown, but
demonstrated many unusual properties that could not be explained within
existing theories.

Here we analyse the experimental data from Ref.~3, as well as new data
obtained with the same method, employing a theoretical model based upon an
idea proposed by Dyakonov~\cite{4}, which we have developed to account for the
specific features of the contactless experiment.

The basic features of the experiment are as follows
(full details are given in Ref.~3). A sample (typical size $\sim 1$ cm) 
is suspended on the filament of a torque magnetometer and
placed into a superconducting solenoid. The magnetic field is then swept at
a constant rate. The component of the magnetic field perpendicular to the
sample induces a circulating electric field that drives eddy currents in
the 2DES. The currents create a magnetic moment which is detected by the
magnetometer.
Sharp peaks in magnetic field dependence of the induced magnetisation
are observed at integer filling factors, $\nu $ (Fig.1). This is a
manifestation of vanishing longitudinal resistivity, $\rho _{xx}$, 
in the integer QHE regime. 
The saturation of the peak height with increasing the sweep rate is observed 
in all the samples. 
This feature is associated with the QHE breakdown: the circulating
electric field fails to induce currents larger than a certain value. The
saturation value of the magnetic moment, $M_{S}$, is the main quantitative
characteristic of the observed effect. 
The experiment reveals a few astounding features which can not be explained
by any previous theory of the QHE breakdown:

1) $M_S$  can reach much higher values in `dirty', more disordered, 
samples with lower mobility of charge carriers at zero magnetic
field.

2) Induced currents can be detected only at fairly low temperatures 
(typically $T<1.5$K), well below the cyclotron energy.

\begin{figure}[h]
\leavevmode
\centerline{\qquad \epsfysize=3.5in \epsfbox{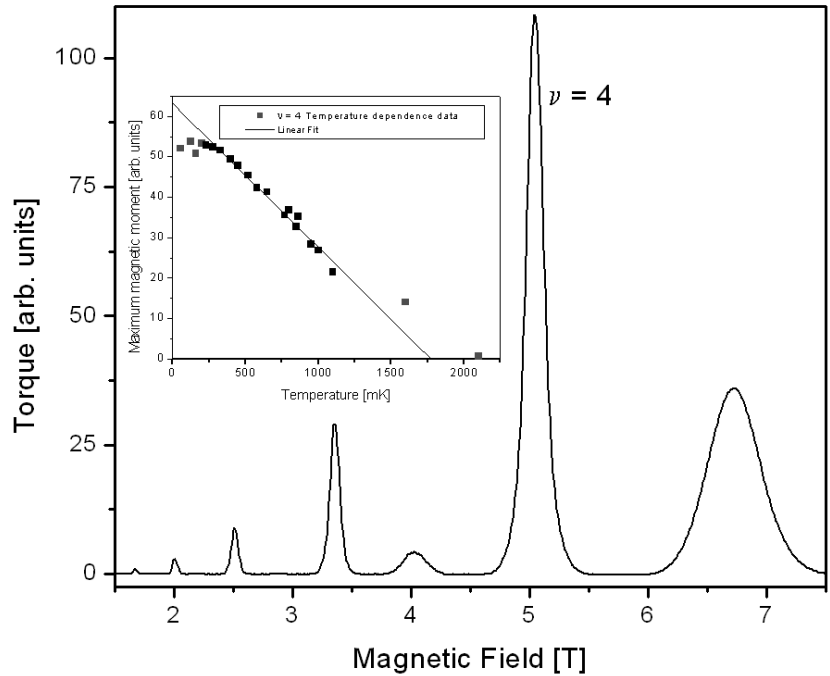}}
\caption{Induced magnetic moment vs magnetic field at a 
constant sweep rate; Inset:
saturation value of the induced magnetic moment vs temperature 
for a `dirty' sample}
\label{Fig1}
\end{figure}

3) In `dirty' samples, the temperature dependence of $M_{S}$ is best
fitted by a descending straight line (see Fig.~\ref{Fig1}, inset); 
in some cases, a `shelf' at low temperatures is observed.

4) All these features are apparently insensitive to the polarity of
charge carriers and the chemical composition of the structure. 
Similar behaviour has been observed for electrons in a GaAs/(Al,Ga)As
heterostructure and for holes in a SiGe heterostructure.
\vspace{5mm}

Our theoretical model considers a disc-shaped sample with the radius $R$. 
The induced magnetic moment $M$ can be written as an integral over the 
sample of the tangential component of the current density, 
$j_{\varphi }(r)$:
\begin{equation}
M=\frac{\pi }{c}\int\limits_{0}^{R}j_{\varphi }(r)r^{2}dr=
\frac{\pi }{c}\int\limits_{0}^{R}\sigma _{xy}E_{r}(r)r^{2}dr,
\label{moment}
\end{equation}
where $E_r$ is the radial component of the in-plane electric field. 
Here we used the fact that in the QHE regime 
$j_{\varphi}\approx \sigma _{xy}E_{r}$.

\begin{figure}[h] 
\leavevmode
\centering{\epsfbox{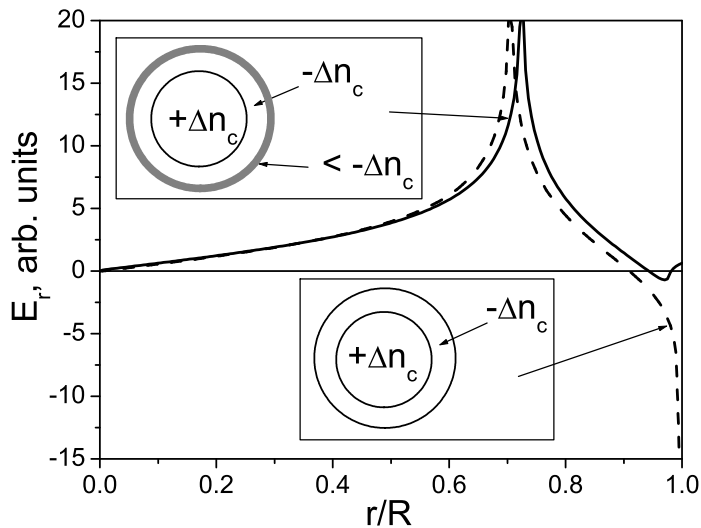}}
\caption{Charge (insets) and radial electric field 
distribution across the sample.}
\label{Fig2}
\end{figure}

Thus, to calculate the magnetic moment one has to find the radial electric
field throughout the sample. This electric field is created by charges which
are redistributed within the sample in such a way that the sample as a whole
remains electrically neutral. As noticed by Dyakonov \cite{4}, 
in two-dimensional
systems these charges cannot be concentrated at the edges, but must be
distributed over the plane. In contactless experiments with disc-shaped
samples, the validity of this statement is evident: there are simply no edge
states in the centre of the sample, where some excess charge (positive or
negative) should be placed. The excess or deficit, 
$\Delta n$, of charge carriers 
in the 2DES plane results in shifting the Fermi-level, 
$\varepsilon _{F}$, from its
initial position ($\varepsilon =0$).
The probability of thermal activation of a charge carrier (an
electron to the lowest empty level, or a hole to the uppermost filled level)
is proportional to $\exp \left[-\left(\varepsilon _{0}/2-\varepsilon
_{F}(\Delta n)\right)/k_{B}T\right] +\exp \left[ -\left(\varepsilon
_{0}/2+\varepsilon _{F}(\Delta n)\right)/k_{B}T\right] $ 
(where $\varepsilon _{0}$
is the energy distance between the uppermost filled and the lowest empty
electron levels in an idealised 2DES). As long as $\left|\Delta n\right| $ 
is small, this
is a very small number. With an increase of the absolute value of 
$\Delta n$, one of the exponentials increases and the other decreases, so 
that it can be neglected. This results in an exponential dependence of the 
longitudinal conductivity, 
$\sigma _{xx}$, on $\left| \varepsilon _{F}(\Delta n)\right| $.
As a consequence, $\sigma _{xx}$ should demonstrate a threshold behaviour:
there is no mobile charge until the exponent is less than some critical
value; above this value, the conductivity is high enough to provide charge
relaxation, so that the radial field, $E_{r}$, can not be sustained. The
threshold condition can be written as:
\begin{equation}
\exp \left( -\frac{\varepsilon _{0}/2-\left| \varepsilon _{F}
(\Delta n)\right| }{k_{B}T}\right) =C, 
\label{Cdefenition}
\end{equation}
where $C$ is a small number. Using Eq.~(2) we find the threshold
value of $\left| \Delta n\right|$ as
\begin{equation}
\Delta n_{c}(T) =
\int\limits_{0}^{\varepsilon _{F}}\rho (\varepsilon )d\varepsilon 
=\int\limits_{0}^{\varepsilon _{0}/2+k_{B}T\ln
C}\rho (\varepsilon )d\varepsilon,   
\label{ncritical}
\end{equation}
where $\rho(\varepsilon)$ is the density of localised electron states. 
This result allows us to construct the electric field distribution
in the sample which would correspond to saturation of the magnetic moment 
in the QHE regime. Indeed, the threshold behavior of conductivity allows 
two situations: (i) $\left| \Delta n \right| <  \Delta n_{c}(T) $ at
any $E_r$ or (ii) $\left| \Delta n\right| > \Delta n_{c}(T)$
at $E_{r}=0$. As we are interested in charge distributions that would 
provide maximum possible $E_{r}$, we construct it on the basis of
condition (i) using, however, condition (ii)
in small parts of the sample to provide self-consistency of the solution. 
The obvious first approximation for the charge distribution which gives 
the maximum possible value of $M$ is the
following: $+e \Delta n_{c}(T)$ at $r<R/\sqrt{2}$ and 
$-e \Delta n_{c}(T)$ at $R/\sqrt{2}<r<R$ (or the same with 
the opposite sign, depending on the direction of the magnetic-field sweep). 
This distribution provides overall neutrality of the sample, while maximum
possible charge is moved. The electric field created by this charge
distribution in the sample plane can be expressed in terms of elliptic
integrals. It is plotted against $r/R$ in Fig.~2 (dashed line). Non-physical
negative values of the electric field near the edge indicate that {\it more}
charge can be placed in this region. Indeed, placing an additional charge
density $-e\delta n=-e\Delta n_{c}(T)\left[\sqrt{b/(R-r)}-1\right]$ 
into a narrow strip of width $b\approx 0.02R$ 
along the edge eliminates the singularity and makes the field in this
region close to zero, in accordance with condition (ii) (solid line in
Fig.~\ref{Fig2}). 
Further refinement of the dependence of $E_{r}$ on $r$ is possible,
but is not needed, since the contribution of the narrow edge region to the
magnetic moment (Eq.~(1)) is negligible. As the charge density over all the
sample, and therefore $E_{r}$, is proportional to 
$ \Delta n_{c}(T)$, the saturation value of the 
magnetic moment $M_{S}$, according to Eq.(1), is
\begin{equation}
M_{S}=\Lambda \frac{\pi }{c}\sigma _{xy}e \Delta n_{c}(T)
 R^{3}
\label{MSfinal}
\end{equation}
where $\Lambda $ is the value of the integral in Eq.~(1) expressed in
dimensionless variables; our numerical calculations give 
$\Lambda \approx 1.1$. 
Thus, the temperature dependence of $M_{S}$ is given by Eq.~(3) for $
\Delta n_{c}(T)$. According to Eq.~(3), $ \Delta n_{c}(T)$
is the number of localised electron states within the energy range from zero
(half-way between the filled and empty levels) to  $\varepsilon
_{0}/2+k_{B}T\ln C$ (note that $\ln C$ is negative). 
$ \Delta n_{c}(T)$ vanishes at
temperature $T_{c}=\varepsilon _{0}/2k_{B}\left| \ln C\right|$. 
As $T\rightarrow 0$,  $ \Delta n_{c}(T)$ 
is one half of the total density of localised
states in between the Landau levels. The behaviour of $ \Delta
n_{c}$ at $T<T_{c}$ is derived from the specific energy dependence
of the density of states. Notably, if $\rho$ is approximately constant
everywhere excepting the vicinity of LLs, as expected for samples with
strong disorder, the temperature dependence of $M_S$  is linear:
\begin{equation}
M_{S}=\Lambda \frac{\pi }{c}\sigma _{xy}eR^{3}
\left(\varepsilon_{0}/2-k_{B}T\left| \ln C\right| \right) \rho   
\label{MSlinear}
\end{equation}
in agreement with experimental results for `dirty' samples. 

One can see from Eq.~(5) that $M_{S}$ is proportional to the density of
localised states. Therefore, the more disordered the sample, the higher
the current densities it can sustain. However, the radial electric
field will eventually reach values comparable to the QUILLS critical field, 
and this new
mechanism will prevent $M_{S}$ from growing further. This should result in
saturation of $M_{S}$ as a function of temperature when $T$ approaches zero.
This effect is indeed observed.  The QUILLS conditions
should first be met near $r=R/\sqrt{2}$, where the electric field is the
highest. The critical fields calculated from our experimental data under 
this assumption agree with the previous theories of QUILLS.


\begin{thebibliography}{4} 
\bibitem{1} G.~Nachtwei, {\em Physica E} {\bf 4}, 
79 (1999) and references therein. 
\bibitem{2} L.~Eaves and F.~W.~Shread, {\em Semicond. Sci. Technol.} {\bf 1}, 
346 (1986).
\bibitem{3} J.P.~Watts, A.~Usher, A.J.~Matthews, M.~Zhu, 
M.~Elliot, W.G.Herrenden-Harker, P.R.~Morris,
M.Y.~Simmons and D.A.~Ritchie, {\em Phys. Rev. Letters} 
{\bf 81}, 4220 (1998);
A.J.~Matthews, J.P.~Watts, M.~Zhu, A.~Usher,
M.~Elliot, W.G.Herrenden-Harker, P.R.~Morris,
M.Y.~Simmons and D.A.~Ritchie, {\em Physica E} {\bf 6}, 140 (2000).
\bibitem{4} M.I.~Dyakonov, {\em Solid State Commun.} {\bf 78}, 817 (1991). 
\end{thebibliography}
\end{document}